# Self-Selected or Mandated, Open Access Increases Citation Impact for Higher Quality Research

**Yassine Gargouri[1], Chawki Hajjem[1], Vincent Larivière[2], Yves Gingras[3], Les Carr[5], Tim Brody[5] & Stevan Harnad,[4,5]**

[1]Institut des sciences cognitives, U. du Québec à Montréal
[2]Observatoire des Sciences et des Technologies, U. du Québec à Montréal
[3]Canada Research Chair in the History and Sociology of Science, U. du Québec à Montréal
[4]Canada Research Chair in Cognitive Sciences, U. du Québec à Montréal
[5]School of Electronics and Computer Science, U. of Southampton
*Corresponding Author*: S. Harnad, Institut des sciences cognitives, Université du Québec à Montréal, Montréal, Québec CANADA H3C 3P8 harnad@uqam.ca

**Abstract:**

**Background:** Articles whose authors make them Open Access (OA) by self-archiving them online are cited significantly more than articles accessible only to subscribers. Some have suggested that this "OA Advantage" may not be causal but just a self-selection bias, because authors preferentially make higher-quality articles OA. To test this we compared self-selective self-archiving with mandatory self-archiving for a sample of 27,197 articles published 2002-2006 in 1,984 journals.

**Methdology/Principal Findings:** The OA Advantage proved just as high for both. Logistic regression showed that the advantage is independent of other correlates of citations (article age; journal impact factor; number of co-authors, references or pages; field; article type; or country) and greatest for the most highly cited articles. The OA Advantage is real, independent and causal, but skewed. Its size is indeed correlated with quality, just as citations themselves are (the top 20% of articles receive about 80% of all citations).

**Conclusions/Significance:** The OA advantage is greater for the more citeable articles, not because of a *quality bias* from authors self-selecting what to make OA, but because of a *quality advantage,* from users self-selecting what to use and cite, freed by OA from the constraints of selective accessibility to subscribers only. It is hoped that these findings will help motivate the adoption of OA mandates be universities, research institutions and research funders.

**One-sentence Summary:** *We demonstrate that the greater citation impact of open access research is causal rather than an artifact of author bias (i.e., authors self-selectively making higher quality research open access) by showing that the citation increase is just as great when the open access is mandatory; the open access impact advantage is independent of other correlates of citation impact, and greater for higher quality research.*

## Introduction

The 25,000 peer-reviewed journals and refereed conference proceedings that exist today publish about 2.5 million articles per year, across all disciplines, languages and nations. No university or research institution anywhere, not even the richest, can afford to subscribe to all or most of the journals that its researchers may need to use (Odlyzko 2006). As a consequence, all articles are currently losing some portion of their potential research impact (usage and citations), because they are not accessible online to all their potential users (Hitchock 2009).

This is supported by recent evidence, independently confirmed by many studies, that articles whose authors have supplemented subscription-based access to the publisher's version by self-archiving their own final draft to make it accessible free for all on the web (“Open Access”, OA) average twice as many citations as articles in the same journal and year that have not been made OA. This “OA Impact Advantage” has been found in all fields analyzed so far -- physical, technological, biological and social sciences, and humanities (Lawrence 2001; Brody & Harnad 2004; Hajjem et al. 2005; Moed 2005b; Eysenbach 2006; Giles et al. 1998; Kurtz & Brody, 2006; Norris et al. 2008; Evans 2008; Evans & Reimer 2009).

Hence OA is not just about public access rights or the general dissemination of knowledge: It is about increasing the impact and thereby the progress of research itself. A work’s research impact is an indication of how much it contributes to further research by other scientists and scholars, how much it is used, applied and built upon (Brin & Page 1998; Garfield 1955, 1976, 1988; Page et al. 1999). That is also why impact is valued, measured and rewarded in researcher performance assessement as well as in research funding (Harnad 2009).

**Self-archiving mandates**

Only about 15% of the 2.5 million articles published annually worldwide are being self-archived by their authors today (Bjork et al 2008; Hajjem and al., 2005). Creating an Institutional Repository (IR) and encouraging faculty to self-archive their articles therein is a good first step, but that is not sufficient to raise the self-archiving rate appreciably above its current spontaneous self-selective baseline of 15% (Sale, 2006). Nor are mere requests or recommendations by researchers’ institutions or funders, encouraging them to self-archive, enough to raise this 15% figure appreciably, even when coupled with offers of help, rewards, incentives and offers to do the deposit on the author’s behalf. In two international, multidisciplinary surveys, 95% of researchers reported that they would self-archive if (but only if) *required* to do so by their institutions or funders. (Eighty-one percent reported that, if it was required, they would deposit *willingly*; 14% said they would deposit reluctantly, and only 5% would not comply with the deposit requirement; Swan 2006.) Subsequent studies on actual mandate compliance have gone on to confirm that researchers do indeed do as they reported they would, with mandated IRs generating deposit rates several times greater than the 15% self-selective baseline and well on the road toward 100% within about two years of adoption (Sale, 2006).

Universities' own IRs are the natural locus for the direct deposit of their own research output: Universities (and research institutions) are the universal providers of all research output, in all scientific and scholarly disciplines; they accordingly have a direct interest in hosting, archiving, monitoring, measuring, managing, evaluating, and showcasing their own research output in their own IRs, as well as in maximizing its uptake, usage, and impact (Holmes & Oppenheim 2001; Oppenheim 1996; . OA self-archiving mandates hence add visibility and value at both the individual and institutional level (Swan & Carr 2008).

In 2002, The University of Southampton's School of Electronics & Computer Science (ECS) became the first in the world to adopt an official self-archiving mandate. Since then, a growing number of departments, faculties and institutions worldwide (including Harvard, Stanford, and MIT) as well as research funders (including all seven UK Research Funding Councils, the US National Institutes of Health, and the European Research Council) have likewise adopted OA self-archiving mandates. Over 100 mandates had already been adopted and registered and charted in ROARMAP[1] as of autumn 2009.

In 2008, mindful of the benefits of mandating OA, the council of the European Universities Association (EUA)[2] unanimously recommended that all European Universities should create IRs and mandate that all their research output should be deposited in them immediately upon publication (to be made OA as soon as possible thereafter). The EUA further recommended that these self-archiving mandates be extended to all research results arising from EU research project funding. A similar recommendation was made by EURAB (European Research Advisory Board). In the US, the FRPAA has proposed similar mandates for all research funded by the major US research funding agencies.

Some studies, however, have suggested that the "OA Advantage" might just be a self-selection bias rather than a causal factor, with authors selectively tending to make higher-quality (hence more citeable) articles OA (Craig et al. 2007; Davis & Fromerth 2007; Henneken et al 2006; Moed 2006). The present study was carried out to test this hypothesis by comparing self-selected OA with mandated OA on the basis of the research article output of the four institutions with the longest-standing OA mandates: (i) Southampton University (School of Electronics & Computer Science) in the UK (since 2002); (ii) CERN (European Organization for Nuclear Research) in Switzerland (since November, 2003); (iii) Queensland University of Technology in Australia (since February 2004); (iv) Minho University in Portugal (since December, 2004).

## Method

The objective was to compare citation counts -- always within the same journal/year -- for OA (O) and non-OA (Ø) articles, comparing the O/Ø citation ratios for OA that was self-selected (S) vs. mandated (M). (The critical comparisons were hence SO/Ø vs. MO/Ø.) The sample covered articles published between 2002 and 2006.[3] The metadata for the articles were collected from the four institutional repositories, as well as from the Thomson-Reuters citation database.[4]

[1] ROARMAP (Registry of Open Access Repository Material Archiving Policies) http://www.eprints.org/openaccess/policysignup/

[2] EUA consists of more than 800 universities, in 46 countries (in January 2009)

[3] About two years need to elapse for the citations from the most recent year to stabilize.

The effect of OA on citation impact cannot be reliably tested by comparing OA and non-OA journals because no two journals have identical subject matter, track-records and quality-standards (nor are there as yet enough established OA journals in most fields). The comparison must hence be between OA and non-OA articles published within the same (non-OA) journals (Harnad and Brody, 2004). For each mandated article, $M_i$, deposited in our four mandated IRs we accordingly collected, as our pool of nonmandated controls for comparison, all articles $N_j$ published in the same journal, volume and year. Our sample of deposited articles from 2002 to 2006 was distributed across 1,984 non-OA journals in the Thomson-Reuters database (***Table 1***).[5]

| | **Journal Count** |
|---|---|
| **2002** | 331 |
| **2003** | 367 |
| **2004** | 415 |
| **2005** | 445 |
| **2006** | 426 |
| **TOTAL** | 1984 |

***Table 1: Journal counts per year***

To reduce our nonmandated comparison sample to a reasonable processing size, we restricted the number of journal/year-matched controls to the 10 $Ø_j$ articles that were semantically closest to their corresponding target $M_i$ (as computed on the basis of shared words in titles, omitting stop words). This tightening of content similarity also made the control articles even more comparable to their targets than using the full spectrum of same-journal content. The total size of the article sample (6215 mandated targets plus their 20982 corresponding controls[6]) from 2002 to 2006 was 27197.

[4] Citation counts were extracted from the Thomson-Reuters database November, 2008.

[5] Based on the Directory of Open Access Journals (DOAJ), 2% of journals indexed by Thomson-Reuters in 2006 were OA journals. Articles from these journals were removed from our pool because for them O/Ø comparisons were not possible.

[6] When more than one M article was published in the same journal/volume/year (which represents 66% of M articles), only 10 articles were selected as controls, using keyword matching for one of these M articles.

The full-text OA status of the articles in our sample was verified using an automated webwide search-robot (Hajjem and al. 2005) as well as an automated Google Scholar search. ***Figure 1*** shows each of our four mandated institutions' verified annual OA article deposits as a percentage of the institution's total published article output for each year based (only) on those articles published in the journals indexed by the Thomson-Reuters citation database; the resulting estimate of the overall OA mandate compliance rate is about 60%. Note also the robot data's confirmation of the ~15% baseline for spontaneous, self-selected (i.e., non-mandated) OA self-archiving among the control articles in the same journal/years.

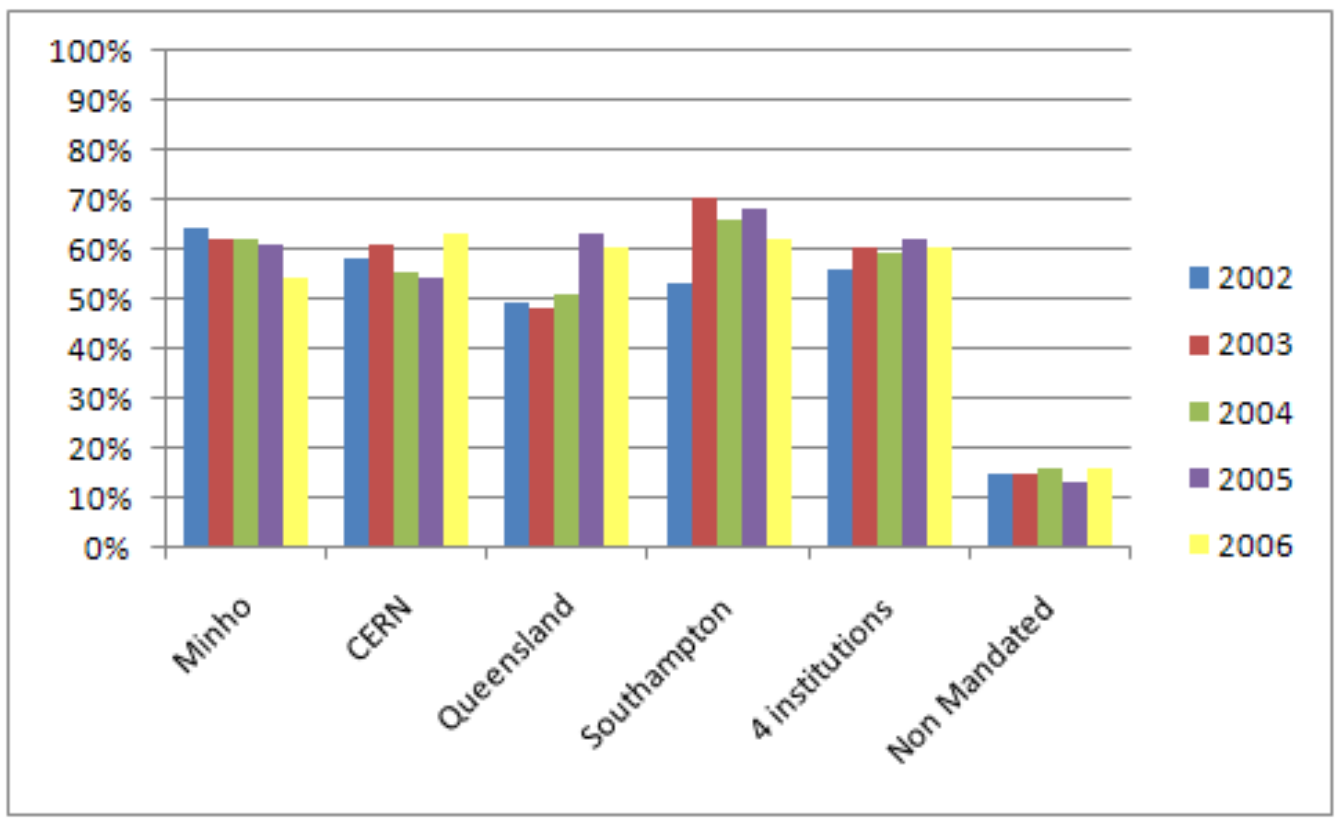

**Figure 1:** ***OA Self-Archiving Levels for Mandate Compliance and Self-Selected Controls:*** As estimated from their Thomson-Reuters-indexed portion, about 60% of each mandated institution's total yearly article output was deposited and hence made OA as mandated. The corresponding percentage OA among the control articles published in the same journal/year (but originating from other, presumably nonmandated institutions) was close to the previously reported global spontaneous baseline rate of about 15% for self-selected (nonmandated) self-archiving (Bjork et al 2008).

This mandated deposit rate of 60% is substantially higher than the self-selected deposit rate of 15%. Although with anything short of 100% compliance it is always logically possible to hold onto the hypothesis that the OA citation advantage could be solely a self-selection bias -- arguing that, with a mandate, the former bias in favor of self-selectively self-archiving one's more citeable articles takes the form of a selective bias against compliance with the mandate for one's less citeable articles), but a reasonable expectation would be at least a diminution in the size of the OA impact advantage with a mandated deposit rate four times as high as the spontaneous rate, if it is indeed true that the OA advantage is solely or largely due to self-selection bias.

To test whether mandated OA diminishes the OA citation advantage, 4 kinds of articles need to be compared:

- O M : OA, Mandated,

- Ø M : Non-OA, Mandated,
- O S : OA, Self-Selected
- Ø S : Non-OA, Self-Selected

The analysis uses the citation counts within each journal/year. Because the date on which the mandate was first adopted varies (from 2002 to 2004) for the four institutions, we analyzed the data for the four institutions separately as well as their joint averages. The separate analyses show the time-course of mandate compliance more clearly; the global analysis combines data, enlarges the sample size and smoothes out incidental effects of institutional and timing differences.

We compare the following ratios: O/Ø, OM/OS, OS/ØS, OM/ØM, OM/Ø, OS/Ø and OM/OS using their mean log citation ratios. For example, to compare mandated OA with self-selected OA, we compute the log of the ratio $OM_j/OS_j$ for each journal *j* and then, we compute the arithmetic mean of all the log ratios for all journals. There is an advantage in favor of OM when the log ratio is greater than 0, and in favor of OS otherwise.

$$OM/OS = \frac{1}{n}\sum_{j=1}^{n}\log\frac{OM_j}{OS_j}$$

The logarithm is used to normalize the data and reduce any effect coming from articles having a relatively high citation count, compared to the whole sample. The comparison is within-journal, to minimize between-journal differences in citation average ("journal impact factor"), and it is keyword-matched to minimize differences still further.

**Results**

Overall, OA articles are cited significantly more than non-OA articles, confirming the repeatedly observed OA Advantage (O/Ø). There is also no evidence at all that mandated OA has a smaller citation advantage than self-selected OA, but rather the contrary. **Figure 2** shows the results for the four institutions together. **Appendix 1** shows each institution separately. The pattern for the individual institutional data is largely the same as for the average across the four institutions.

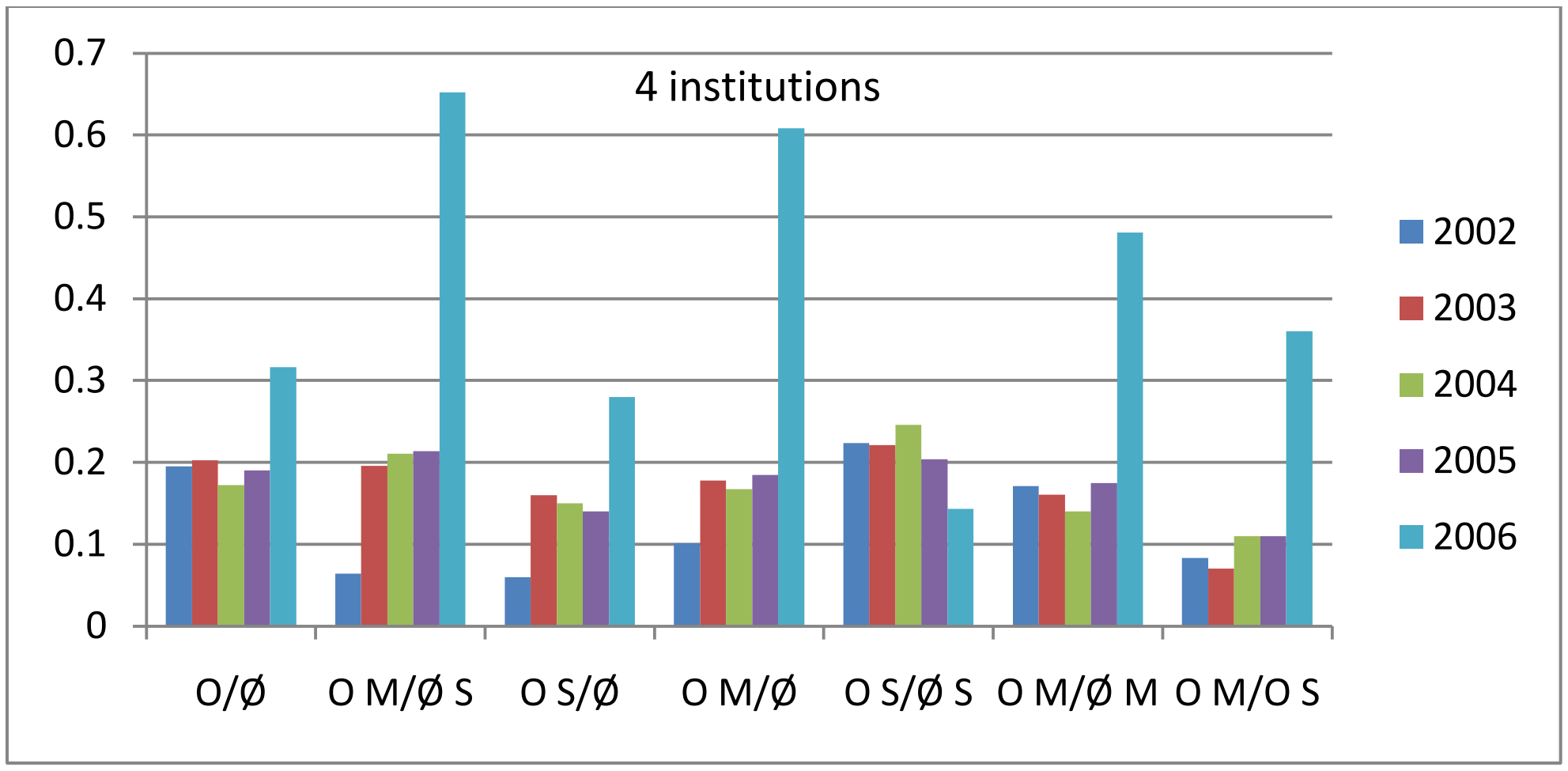

**Figure 2:** ***Comparing the OA Impact Advantage for Self-Selected vs Mandatory OA.*** Averages across the sample of four institutions confirm the significantly higher citation rates for OA vs. matched control Non-OA articles (O/Ø) published in the same journal and year. OA articles are more highly cited irrespective of whether the OA is Self-Selected (S) or Mandated (M). The O/Ø Advantage is present for mandated OA (OM/ØS), and of about the same magnitude, whether S vs M are compared on the basis of the entire control sample (OS/Ø vs OM/Ø) or just S alone vs M alone (OS/ØS vs OM/ØM). Far from the OA Advantage diminishing when it is mandatory (compliance rate, 60%) rather than self-selective (15%), it is, according to this sample, if anything, slightly greater (OM/OS) (although this, being the smallest effect, might be due to chance or sampling error).

**Table 2:** ***Paired Samples Test***

| | | Paired Differences | | | | | | | |
|---|---|---|---|---|---|---|---|---|---|
| | | | | | 95% Confidence Interval of the Difference | | | | |
| | | Mean | Std. Deviation | Std. Error Mean | Lower | Upper | t | df | Sig. (2-tailed) |
| Pair 1 | O - Ø | 1,281651 | 7,401508 | ,250074 | ,790837 | 1,772466 | 5,125 | 875 | ,000 |
| Pair 2 | O_M - O_N | ,834114 | 7,897669 | ,420947 | ,006218 | 1,662010 | 1,982 | 351 | ,048 |
| Pair 3 | O_N - Ø_N | ,808574 | 7,535069 | ,297385 | ,224606 | 1,392541 | 2,719 | 641 | ,007 |
| Pair 4 | O_M - Ø_M | 1,539534 | 9,525721 | ,683907 | ,190643 | 2,888425 | 2,251 | 193 | ,026 |
| Pair 5 | O_M - Ø | ,705398 | 6,871467 | ,285568 | ,144520 | 1,266276 | 2,470 | 578 | ,014 |
| Pair 6 | O_N - Ø | ,890460 | 7,475119 | ,294561 | ,312042 | 1,468878 | 3,023 | 643 | ,003 |
| Pair 7 | O_M - Ø_N | ,646946 | 6,920703 | ,288613 | ,080079 | 1,213812 | 2,242 | 574 | ,025 |

For all OA vs Non-OA (O/Ø) comparisons, regardless of whether the OA was Self-Selected (S) or Mandated (M), the mean log citation differences are significantly greater than 0 (based on correlated-sample t-tests for within-journal differences; **Table 2**). As the last of the four institutional mandates was adopted in 2004, the test was based on a sample of M and S articles published between 2004 and 2006[7]. There is no detectable reduction in the size of the OA Advantage for Mandated OA (60%) compared to Self-Selected OA (15%). (The Mandated OA Advantage is, if anything, greater, and also grows with time.) It would require a very complicated argument indeed ("self-selective noncompliance for less citeable articles") to resurrect the hypothesis that the OA Advantage is only or mostly a self-selection bias in the face of these findings. (Such an argument does remain a logical possibility until there is 100% mandate compliance, but an increasingly implausible one.)

## Logistic regression

The number of citations an article receives can be correlated with and hence influenced by a variety of variables. Those variables, in turn, could create another kind of bias. For example, older articles tend to have more citations than younger articles simply because there has been more time to cite them. If OA articles tended to be older than non-OA articles, then article age, rather than OA, could be the cause of the OA Advantage. A way to test whether correlates of citation other than OA are responsible for the OA Advantage is to perform a logistic regression analysis to see whether OA alone is still significantly correlated with higher citations when the correlation with other variables has been partialled out. We have accordingly analyzed the following set of variables potentially influencing citations. Variables 1-8 are known to be correlated with citation counts.

[7] The greater OA Advantage for 2006 might be due to a variety of factors that will be analyzed in future more detailed studies over a longer time base (and taking deposit date into account, alongside publication date and the date at which citations are counted):

(a) These results were analyzed in 2009, and 2006 was the most recent full year analyzed. If the results for a year are analyzed before at least 1.5 years have elapsed, citations are still incomplete, the data are unstable, and the OA Advantage may not yet be detectable.

(b) Some of the compliance rates for 2002-2006 may have been retroactive, with the older articles deposited several years after they were published. That would mean that older articles were not receiving their full OA Advantage (and the finding of an Early Access Advantage by Kurtz et al 2005 and Kurtz & Henneken 2007 suggests that later deposits may never gain all the citations they would have received if deposited earlier). The mandates were adopted between 2003 and 2005. So perhaps only 2006 was receiving its full OA Advantage.

(c) Citations grow with article age but our multiple regression analyses also reveal an OA*Age interaction, with the OA Advantage growing faster than article age. Hence the OA Advantage becomes bigger for older articles, when measured independently, with other variables that increase citations (such as age, journal impact factor, number of co-authors) partialled out.

(d) In contrast to (c), however, it is also possible that as global OA is growing, global OA use is rising, which would mean that the OA Advantage itself is growing; this too could help explain the higher Advantage for the most recent year in our current sample (2006).

Variable 9 is OA itself; and variable 10 is a measure of the degree to which the relation between OA and Age is non-additive. Variable 11 is whether or not the OA is mandated. Variables 12-15 are just the four mandating institutions that are our reference points in this study.

1. **Age** : How old is the article (articles published from 2002 to 2006)?
2. **JIF** : What is the Thomson-Reuters "Impact Factor" (average citations per article in 2-year window) of the journal in which the article was published (from 0 to 30)?
3. **Auth_N** : How many co-authors does the article have?
4. **Ref_N** : How many references does the article cite?
5. **Page_N** : How many pages in the article?
6. **Sci** : Is the article classified by Thomson-Reuters as Science (1) or Social Science (0)
7. **Review** : Is the article classified by Thomson-Reuters as a "review" article (1) or not (0)?
8. **USA** : What is the country of the first author (USA 1, other 0)?
9. **OA**: Is the article Open Access (1) or Not (0)?
10. **Age*OA** : The interaction between Age and OA
11. **M**: Does the author's institution Mandate Open Access (1) or Not (0)?
12. **CERN** : Is the first author from CERN (1/0)?
13. **South** : Is the first author from Southampton (1/0)?
14. **Minho** : Is the first author from Minho (1/0)?
15. **Queens** : Is the first author from Queensland University of Technology (1/0)?

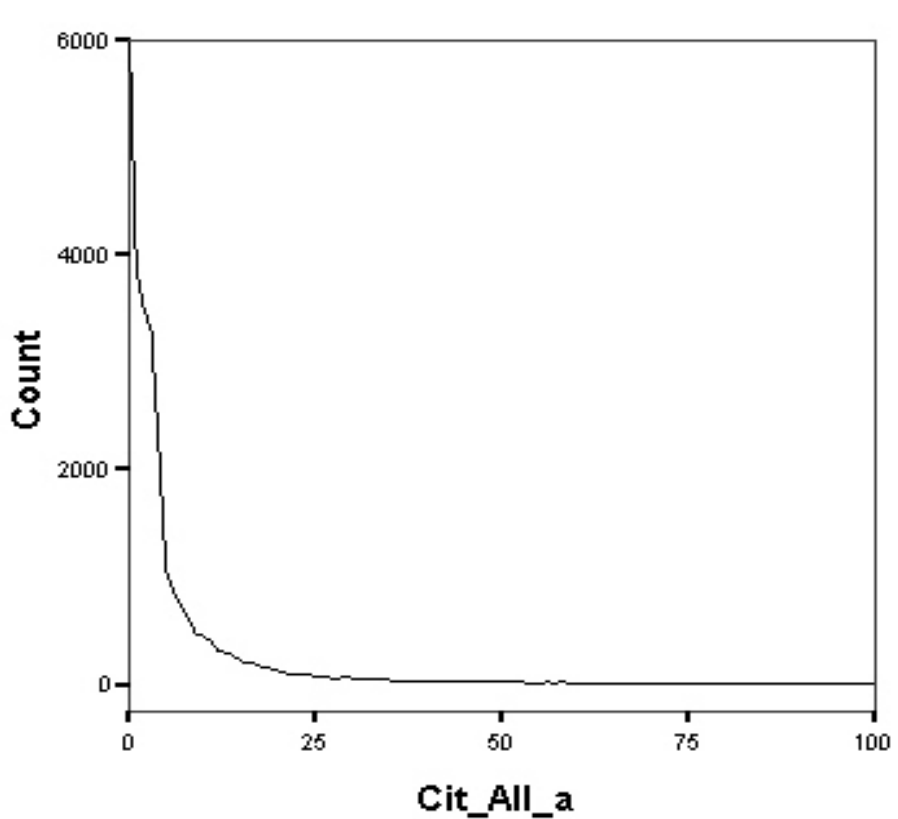

**Figure 3*: Citation count distribution (minus self-citations*)**. 23% of our sample of 27197 articles had zero citations; 51% had 1-5 citations; 12% had 6-10 citations; 8% had 11-20 citations; 6% had 20+ citations.

| Model N. | Dependent V. | Age | JIF | Ref_N | Auth_N | Page_N | OA | M | USA | Review | Sci | CERN | South | Minho | Queens | Age*OA |
|---|---|---|---|---|---|---|---|---|---|---|---|---|---|---|---|---|
| M_1 | Cit_a_0&1-5 | **1.494** | **2.229** | **1.020** | **1.007** | **0.993** | *0.957* | | | **0.627** | **1.249** | **0.789** | | | **1.476** | **1.209** |
| M_2 | Cit_a_1-5&5-10 | **1.490** | **1.514** | **1.016** | **1.002** | **0.986** | **1.323** | **1.889** | **1.415** | **0.777** | **1.475** | | | | | |
| M_3 | Cit_a_1-5&10-20 | **1.786** | **1.776** | **1.020** | *1.002* | **0.992** | **1.392** | **1.716** | **1.406** | *0.992* | **1.887** | | | | | |
| M_4 | Cit_a_1-5&20+ | **2.439** | **2.114** | **1.019** | **0.999** | | **8.953** | | **1.860** | **1.914** | **3.050** | **2.306** | | | | *0.968* |

***Table 3: The Exp(ß) values for logistic regressions***

Bold: p<0.01
Italic: 0.01≤p<0.05

All self-citations were subtracted from the citation counts. (About 32% of the articles in our sample have at least 1 self-citation, with an average of about 2 self-citations per article.) As is well-known and evident from **Figure 3**, citation counts are not normally distributed and instead follow a power-law or stretched-exponential function (Lariviere et al. 2009; Wallace et al. 2009). We accordingly used binary stepwise logistic regression analysis, with a dichotomous dependent variable, selecting for each test the model that maximizes the chi-square likelihood ratio. To make the interpretation of the coefficients easier, we exponentiated the ß coefficients (**Exp(ß)**) and interpreted them as odds-ratios. For example, we can say for the first model that for a one unit increase in OA, the odds of receiving 1-5 citations (versus zero citations) increased by a factor of 0.957. **Table 3** and **Figure 4** show Exp(ß) values for each model having "**Cit_a_x-y&y-z**" as dependent variables ((x,y,z)∈ {1, 2, 3, ..., 20}), where Cit_a_x-y&y-z = 1 if the citation count (minus self-citations) is between y and z and 0 if between x and y. Models are referred to as "**M_r**". (The Exp(ß) values of variables turned out to have the same polarity and to be quite similar, with and without self-citations).

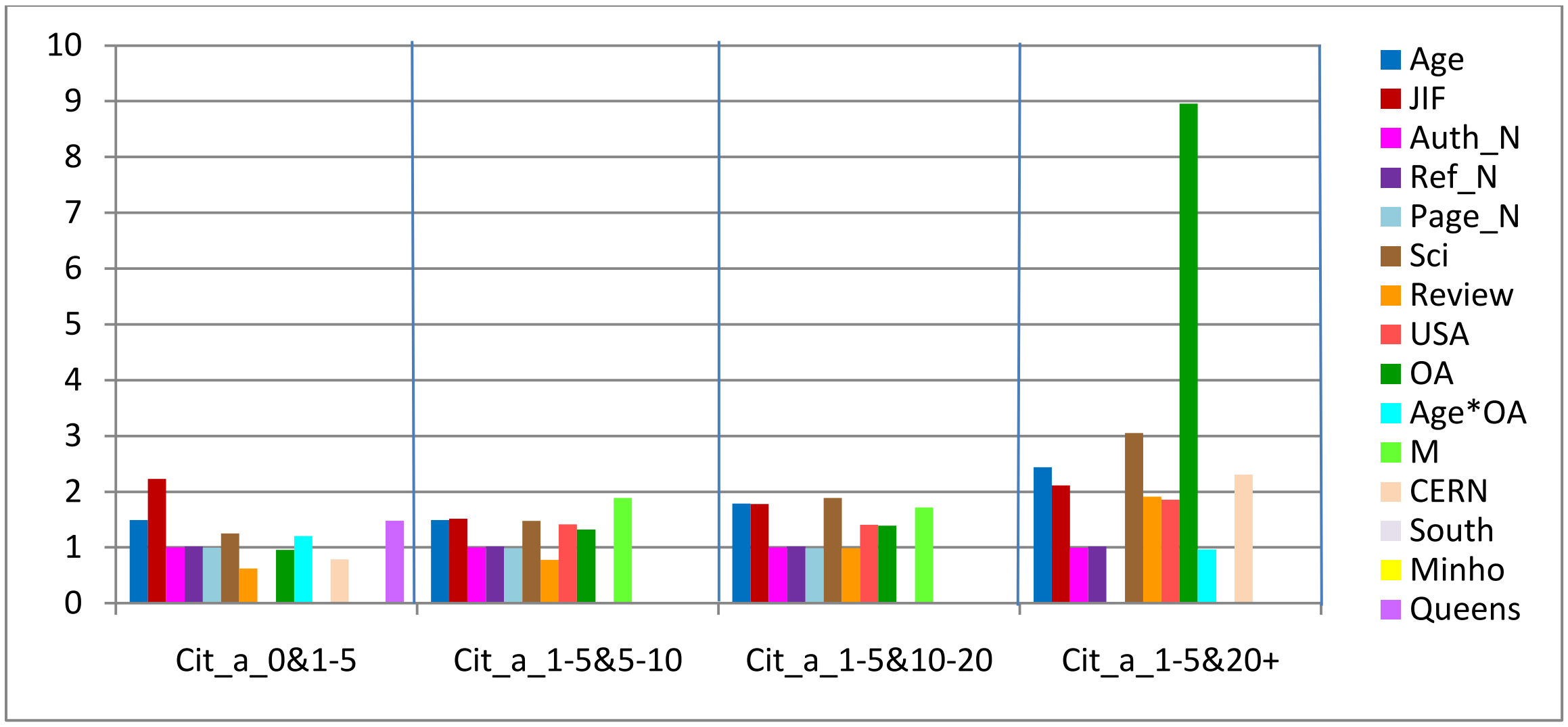

**Figure 4:** ***The Exp(ß) values for logistic regressions.*** In most of the four citation range comparisons (zero/low, low/med1, low/med2, low/high) citation counts are positively correlated with Age, Journal Impact Factor, Number of Authors, Number of References, Number of Pages, Science, Review, USA Author, OA, and Mandatedness. There is also an OA*Age interaction in the top and bottom range. (Citations grow with time; for age-matched articles, the OA Advantage grows even faster with time; **Figure 5**). OA is a significant independent contributor in every citation range, but especially at the high end.

**Figure 4** shows that citations are, as already known, positively correlated with the first eight variables listed earlier (age, journal impact factor, authors, references, pages, science, review, USA) as well as with OA. Articles that are made OA have significantly higher citation counts. This significant OA advantage, which this analysis now shows to be independent of the other variables, is present in every citation range but highest in the highest citation range (1-5 citations vs 20+ citations): In other words, the OA advantage is greatest for highly cited articles.

In our sample, articles by authors at institutions that have OA Mandates have higher citation counts; this effect is present only in the medium-high citation ranges (and is of course also influenced by the level of author compliance with the institutional Mandate, discussed further below). CERN articles have higher citation counts in the lowest and especially the highest citation range. However, when all CERN articles are excluded from our sample, there is no significant change in the other variables.

There is a significant interaction between Age and OA (Age*OA) for the lowest citation range (0 vs. 1-5 citations) as well for highest citation range (1-5 citations vs. 20 citations and more). Both the linear main effect of age and OA, and this nonlinear interaction are statistically significant. **Figure 5** shows the citation mean (**Cit_a_1-5&20**+) for OA and NOA articles corresponding to each Age value. This figure confirms the OA advantage and shows the interaction with age: The OA Advantage becomes even bigger for older articles.

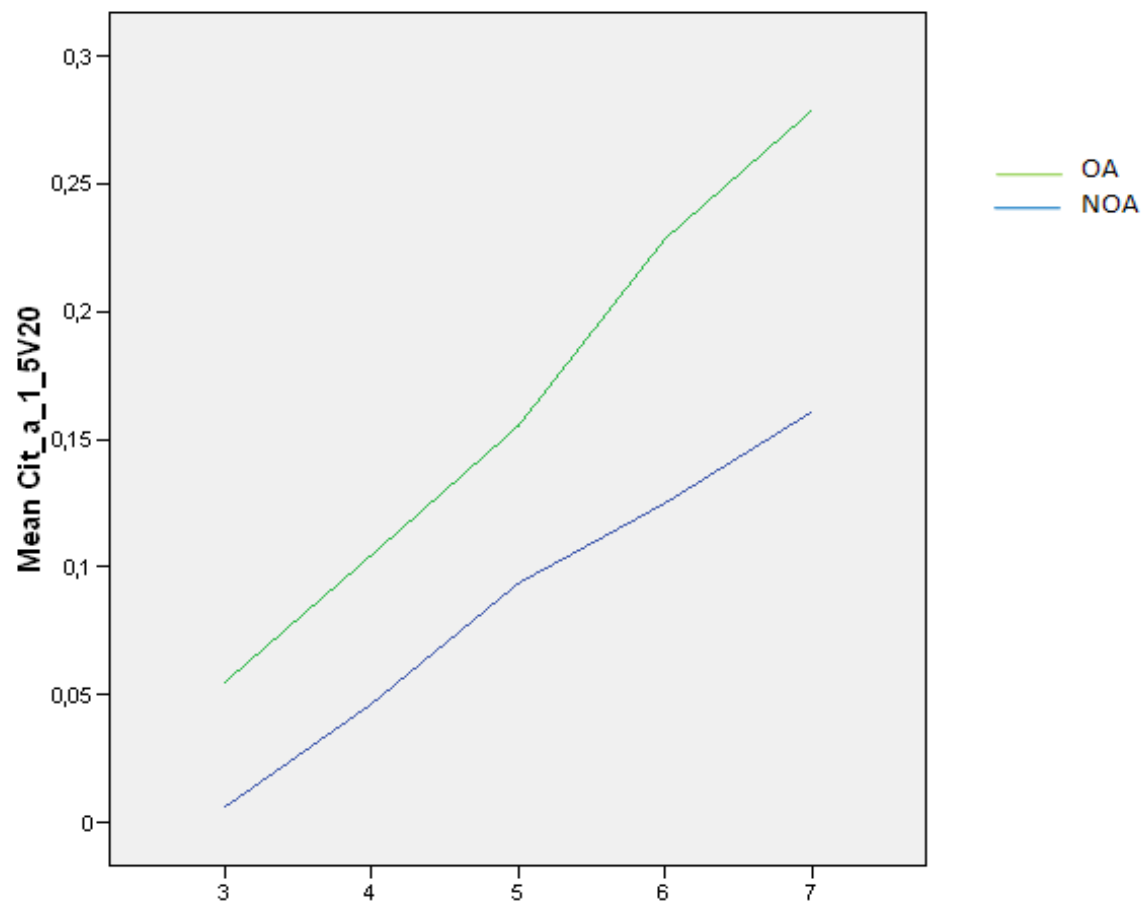

**Figure 5*: Interaction Between Article Age and OA.*** The size of the OA Advantage increases as articles get older, over and above the sum of the independent positive effects of age alone and of OA alone on citations.

**Logistic regression by Impact Factor interval:**

In order to compare articles belonging to comparable journals and to see the profile for journals in increasing impact ranges (see distribution, **Figure 6**), we divided our sample into 4 quartiles by Journal Impact Factor (JIF), each range covering 25% of the articles:

JIF_1 : $0 \leq JIF < 0.633$

JIF_2 : $0.633 \leq JIF < 1.053$Tav

JIF_3 : $1.035 \leq JIF < 1.782$

JIF_4 : $1.782 \leq JIF < 29.957$

Only the top quartile contains journals with JIFs from 1.782 to 29.957. As we are also interested in the variability within this quartile, we further subdivided this top quartile into two octiles, each covering 12.5% of the articles. Subdividing more minutely would make the sample sizes too small to detect effects of interest. This yielded a total of five ranges for the JIF variable:

JIF_1 : $0 \leq JIF < 0.633$

JIF_2 : $0.633 \leq JIF < 1.053$

JIF_3 : $1.053 \leq JIF < 1.782$

JIF_4 : $1.782 \leq JIF < 2.468$

JIF_5 : $2.468 \leq JIF \leq 29.957$

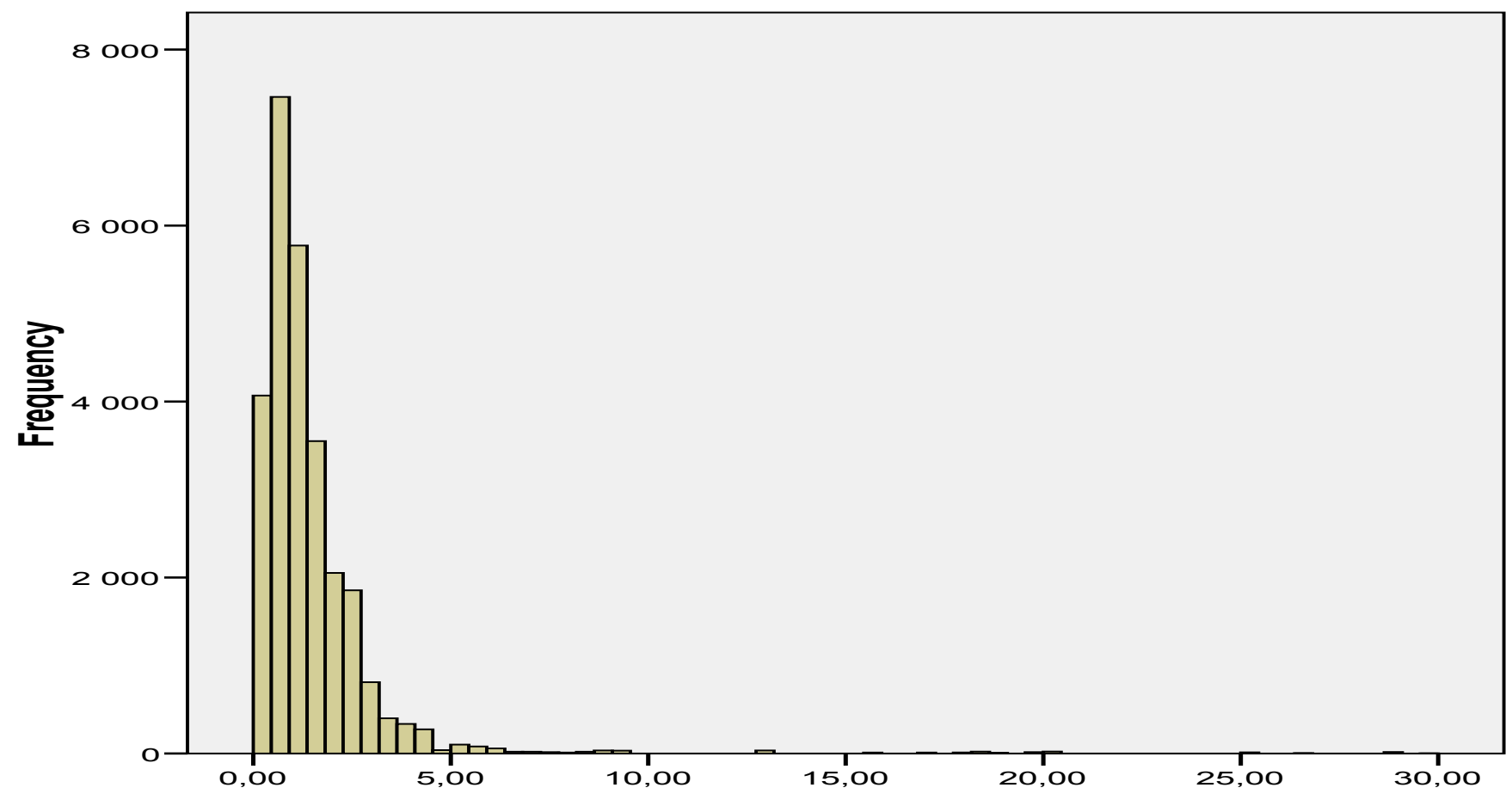

**Figure 6 : *Distribution of journals by Journal Impact Factor (JIF).***

The same regression is done separately for each JIF range by controlling all the variables (except JIF). **Figures 7-11** (and Appendix **Tables 4-8**) summarize the values of **Exp(ß)** corresponding to the controlled variables for each JIF range[8].

When articles are published in a low JIF journal, citation counts for their individual articles are positively correlated with Age, References, Authors, OA and M. The OA advantage is greater in the higher citation ranges. For the lowest range of individual article citations, the Age*OA interaction is significant, but OA itself is not.

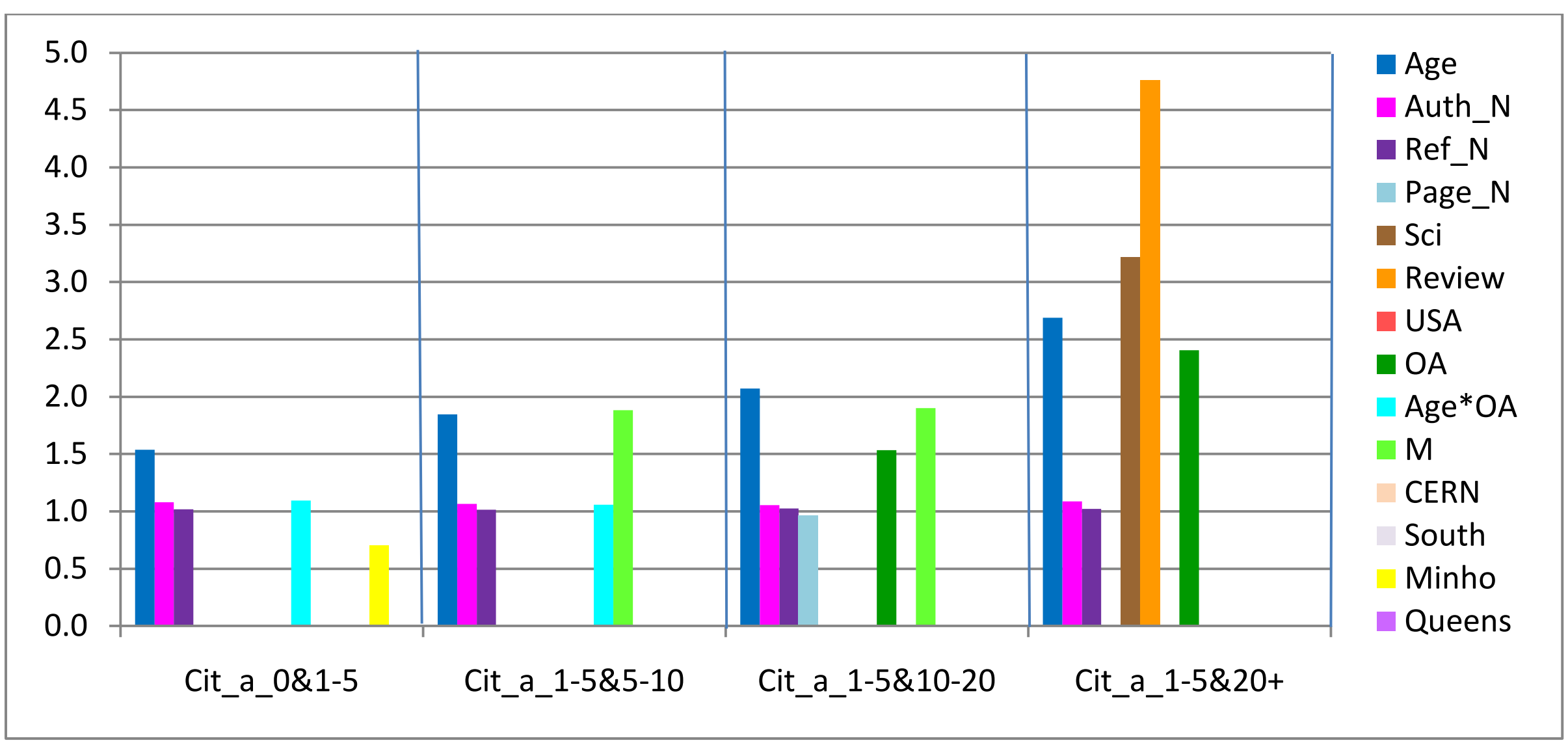

[8] As noted earlier, our Exp(ß) values for these variables have the same polarity and pattern whether or not we exclude self-citations from the citation count.

**Figure 7:** ***Exp(ß) values for logistic regressions (Lowest JIF Range: 0.0-.0.633)***

For articles in journals with JIFs between 0.633 and 1.053, the pattern is quite similar, except the Age*OA interaction is absent and OA itself (alongside Age, as separate variables) is significant.

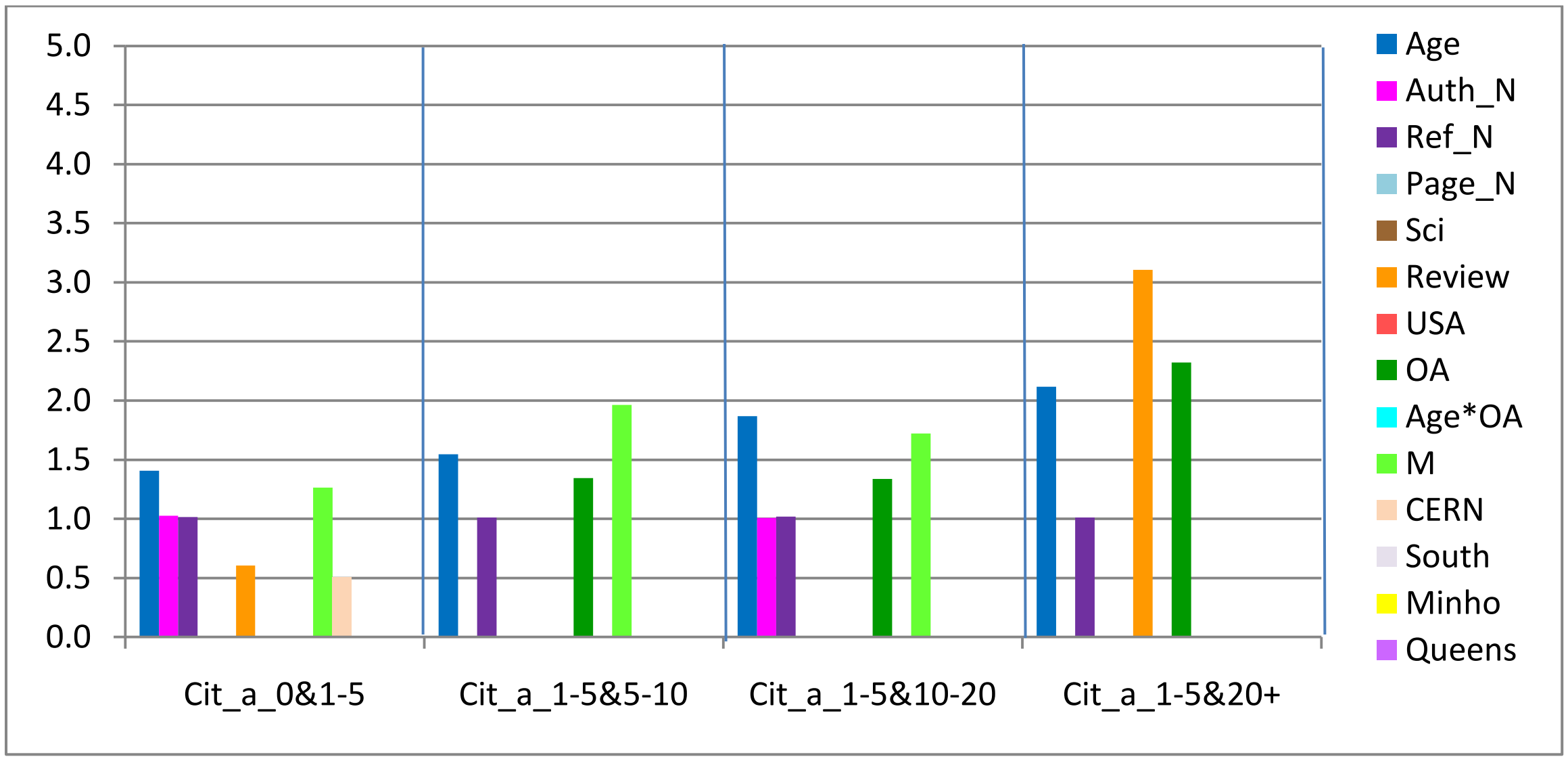

**Figure 8:** ***Exp(ß) values for logistic regressions (JIF range 0.633-1.053)***

For articles in journals with JIFs between 1.053 and 1.782, the pattern is again quite similar. The USA and Review variables now also correlate with citation increase. In this JIF range, three of the institutions (QUT, Southampton and CERN) have a small citation advantage in some of the comparisons. Removing the articles from any one of these institutions, however, does not change the pattern for the other variables.

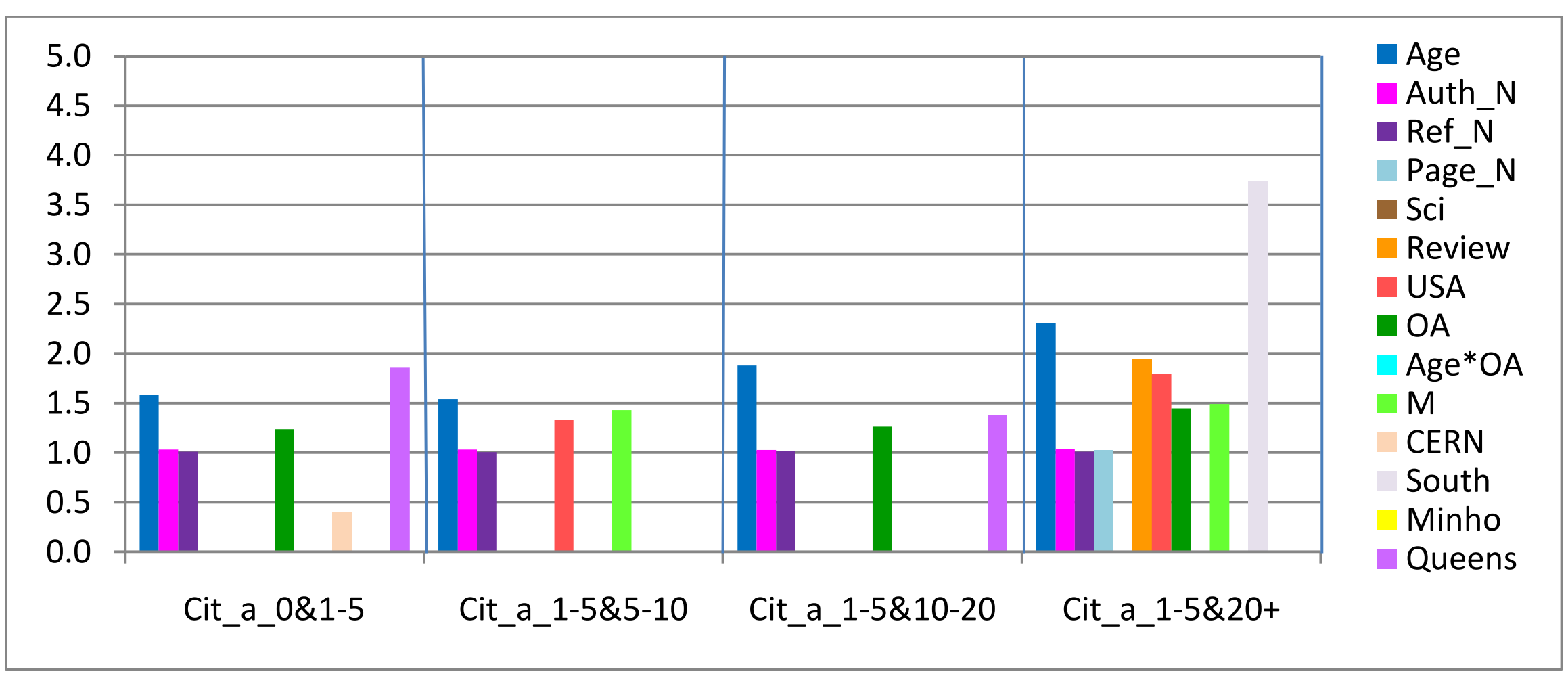

**Figure 9:** ***Exp(ß) values for logistic regressions (JIF range 1.053-1.782)***

For journals with JIFs between 1.782 and 2.468, longer articles (more pages) have more citations. Here the OA advantage is significant only in the highest citation count ranges. The number of authors is also less correlated with increased citations as the citation range gets higher. CERN has a citation advantage in this JIF range. However, removing CERN articles does not alter the pattern for the other variables.

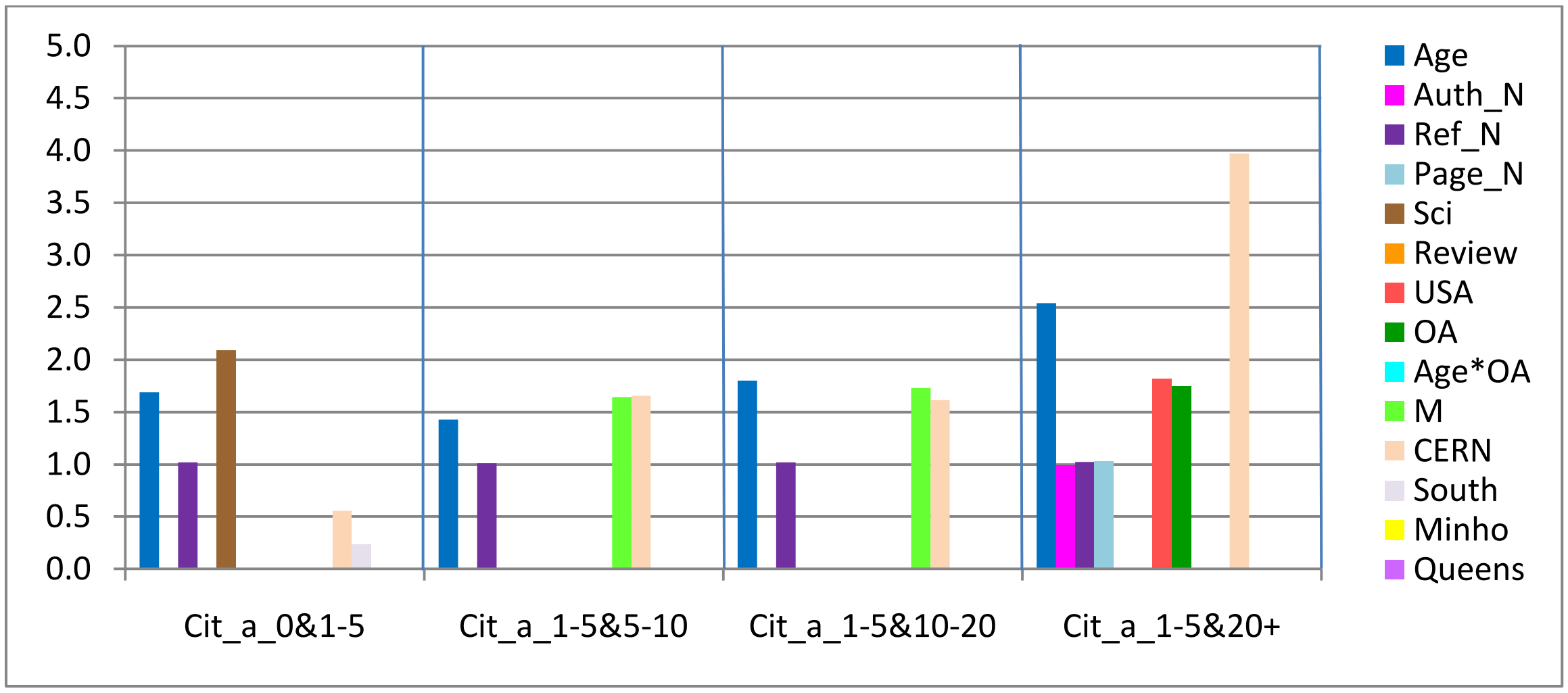

**Figure 10:** ***Exp(ß) values for logistic regressions (JIF range 1.782-2.468)***

For journals with JIFs between 2.468 and 29.957. The OA advantage is again significant for the highest citation ranges. (The increased citations for USA and Review articles also increase in significance.)

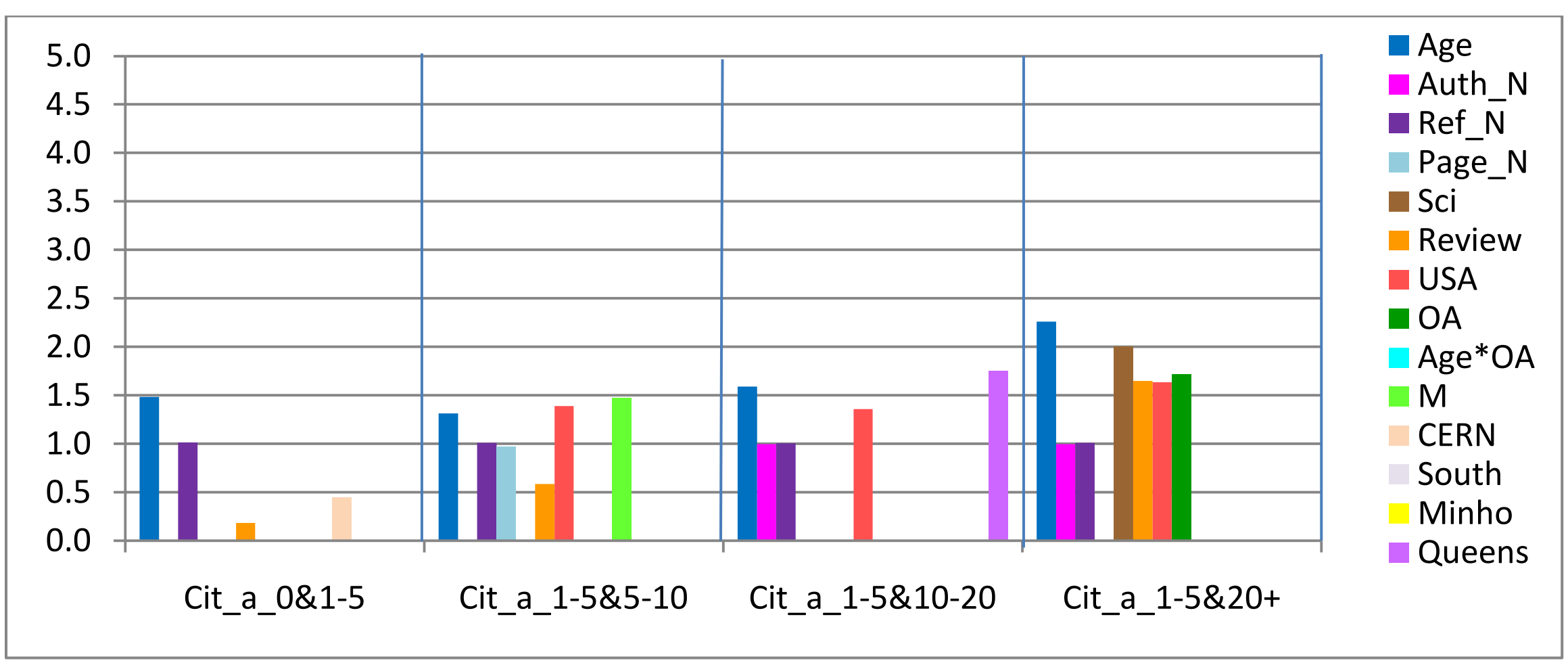

**Figure 11:** ***The Exp(ß) values for logistic regressions (JIF 2.468-29.957)***

Overall, OA is correlated with a significant citation advantage for all journal JIF intervals as well as for the sample as a whole. This advantage is greatest for the highest citation ranges. When regressions are done separately for the different JIF ranges, the Age*OA interaction disappears, but OA and Age (as separate variables) remain significant. (There is no significant effect of a specific institution compared to the rest of the institutions, hence there is no need to exclude any specific institution from our sample.)

**Discussion**

This study confirms that the OA advantage is a statistically significant, independent positive increase in citations, even when we control the independent contributions of many other salient variables (article age, journal impact factor, number of authors, number of pages, number of references cited, Review, Science, USA author). All these other variables are of course correlated with citation counts, so the fact that OA continues to correlate with an independent positive increase in citation counts even when all these other correlates are partialled out is quite a strong outcome. It means that the OA Advantage is not just a bias arising from either a chance or systematic imbalance in the other correlates of citations.

Moreover, the OA advantage is just as great when the OA is mandated (with mandate compliance rate ~60%) as when it is self-selective (self-selection rate ~15%). That makes it highly unlikely that the OA advantage is either entirely or mostly the result of an author bias toward selectively self-archiving higher quality – hence higher citeability – articles.  Nor are the main effects the result of institutional citation advantages, as the institutions were among the independent predictor variables partialled out in the logistic regression; the outcome pattern and significance is also unaltered by removing CERN, the only one of the four institutions that might conceivably have biased the outcome because its papers were all in one field and tended to be of higher quality, hence higher citeability overall.

Since, with the exception of our one unidisciplinary institute,  CERN (high energy physics), the pluridisciplinary articles from the three other mandated institutional repositories are mostly not in fields that habitually self-archive their unrefereed preprints well before publication (as many in high energy physics do), nor in fields that already have effective OA for their published postprints (as astronomy does: Henneken et al 2006, 2008; Kurtz & Brody 2006), it is also unlikely that the OA advantage is either entirely or mostly just an early (prepublication) access advantage (Kurtz et al 2005; Kurtz & Henneken 2007). This will eventually be testable once there are enough reliable deposit-date data, relative to publication-date, for a large enough body of self-archived OA articles. In any case, an early-access advantage in a preprint self-archiving field translates into a generic postpublication OA advantage in fields in which authors do not self-archive prepublication preprints and published postprints are otherwise accessible only to subscribers. The OA mandates all apply only to refereed postprints, not to pre-refereeing preprints.

This study confirms that the OA advantage is substantially greater for articles that have successfully met the quality standards of higher-impact journals and it is also greater in the higher-citation ranges for individual papers within each journal-impact level. The Seglen (1992) "skewness" effect, whereby the top 10-20% of articles receive about 80-90% of all citations, is confirmed in our own sample of 708,219 articles extracted from Thomson-Reuters from 1998 to 2007: In our sample, 20% of *articles* received 80% of all citations (29% received 90%). In addition, 10% of *journals* receive 90% of all citations

The implication is that OA will not make an unuseable (hence unciteable) paper more used and cited (although the proportion of uncited papers has been diminishing with time; Wallace et al. 2009). But wherever there are subscription-based constraints on accessibility, providing OA will increase the usage and citation of the more useable and citeable papers, probably in proportion to their importance and quality, hence citeability. The most likely cause of the OA citation advantage is accordingly not *author self-selection* toward making more citeable articles OA, but *user self-selection* toward using and citing the more citeable articles – once OA self-archiving has made them accessible to all users, rather than just to those whose institutions could afford subscription access. In other words, the OA advantage is a *quality advantage*, rather than a *quality bias*: it is not that the higher quality articles – the ones that are more likely to be selectively cited anyway -- are more likely to be made OA self-selectively by their authors, but that the higher quality articles that are more likely to be selectively cited are made more accessible, hence more citeable, by being made OA.

Our results also suggest that mandated OA might have some further independent citation advantage of its own, over self-selected OA -- but until replicated, it is more likely that this small, previously unreported effect was an effect of chance or sampling error. If there does indeed prove to be an independent "mandate advantage" over and above OA itself, a possible interpretation would be the reverse of the self-selection hypothesis: There may be a higher proportion of higher-quality work among the 85% that are not made OA on a self-selective basis today than among the 15% that are, so the mandates serve to help bring this "cream of science" to the top.

It also needs to be noted that some of the factors contributing to the OA advantage are permanent, whereas others will shrink as OA rises from its current 15% level and will disappear at 100% OA. All *competitive advantage* of OA over non-OA (because OA is more accessible) will of course vanish at 100% OA (as will the possibility of concurrent measurement of the OA Advantage). Any *self-selective bias* (whether positive or negative) too will disappear at 100% OA. What will remain will be the *quality advantage* itself (the tendency of researchers to selectively use and cite the best research, if they can access it), now maximized by making everything accessible to every user online.

There will continue to be the early-access advantage in fast turnaround fields: It is not that making findings accessible earlier merely gets them their citation "quota" earlier; it significantly increases that quota, probably by both accelerating and broadening their uptake in further research (Kutz et al. 2005). And even after the competitive advantage is gone because all articles are OA, the *download advantage* will continue to be enjoyed by all articles (Bollen et al 2009; Davis et al. 2008), while the quality advantage will see to it that for the best work, increased downloads are translated into uptake, usage and eventual increased citations, of which earlier download increases have been found to be predictive (Brody et al. 2006).

**Summary and Conclusion**

The assumption that increasing access to research will increase its usage and impact is the main rationale for the worldwide OA movement. Many prior studies have by now shown across all fields that those journal articles whose authors have made them Open Access by self-archiving them online, freely accessible to all potential users, are cited significantly more than articles that are accessible only to subscribers. There is prior evidence for a self-selection bias in a few special fields (such as astronomy and some areas of physics) where most articles are made OA in unrefereed preprint form long before they are refereed and published, and where the published version is effectively accessible to all potential users as soon as it is published. Authors may indeed be more reluctant to make articles about which they have doubts OA before they are refereed (Kurtz et al. 2005; Moed 2006). We have now shown that for most other fields (i) the OA Advantage remains just as high for mandatory self-archiving as for self-selected self-archiving and that (ii) this is not an artifact of biases in other correlates of citation counts. Both the self-archiving and the mandates apply to refereed postprints, upon acceptance for publication, not to unrefereed preprints.

Hence the OA Advantage is real, independent and causal. It is indeed true that the size of the advantage is correlated with quality, just as citations themselves are (the top 20% of articles receiving about 80% of all citations); but what that means is that the OA advantage is higher for the more citeable articles, not because of a quality bias from author self-selection but because of a quality advantage that OA *enhances* by maximizing accessibility, and thereby also citeability. On a playing field leveled by OA, users can selectively access, use and cite those articles that they judge to be of the highest relevance and quality, no longer constrained by their accessibility.

Overall, only about 15% of articles are being spontaneously self-archived today, self-selectively. To reach 100% OA globally, researchers' institutions and funders need to mandate self-archiving, as they are now increasingly beginning to do. We hope that this demonstration that the OA Impact Advantage is real and causal will give further incentive and impetus for the adoption of OA mandates worldwide in order to allow research to achieve its full impact potential, no longer constrained by limits on accessibility (Brody et al. 2007; Bernius & Hanauske 2009; Carr & Harnad 2009; Dror & Harnad 2009).

To measure that maximized research impact, we and others are already developing new OA metrics for monitoring, analyzing, evaluating, crediting and rewarding research productivity and progress (Adler & Harzing 2009; Bollen et al 2009; Brody 2003; Brody et al 2006; Cronin 1984; Cronin & Meho 2006; De Bellis 2009; De Robbio 2009; Diamond 1986; Harzing 2008; Harnad 2009; Jacso 2006; Moed 2005a). Hence there is no need to have any penalties or sanctions for non-compliance with OA self-archiving mandates. As the experience of Southampton ECS, Minho, QUT and CERN has already demonstrated, OA mandates, together with OA's own rewards (enhanced research access, usage and impact), will be enough to establish the causal connection between providing access and reaping its impact, through the research community's existing system for evaluating and rewarding research productivity. In the online era, researchers' own "mandate" will no longer just be "publish-or-perish" but "self-archive to flourish."

*Appendix 1*

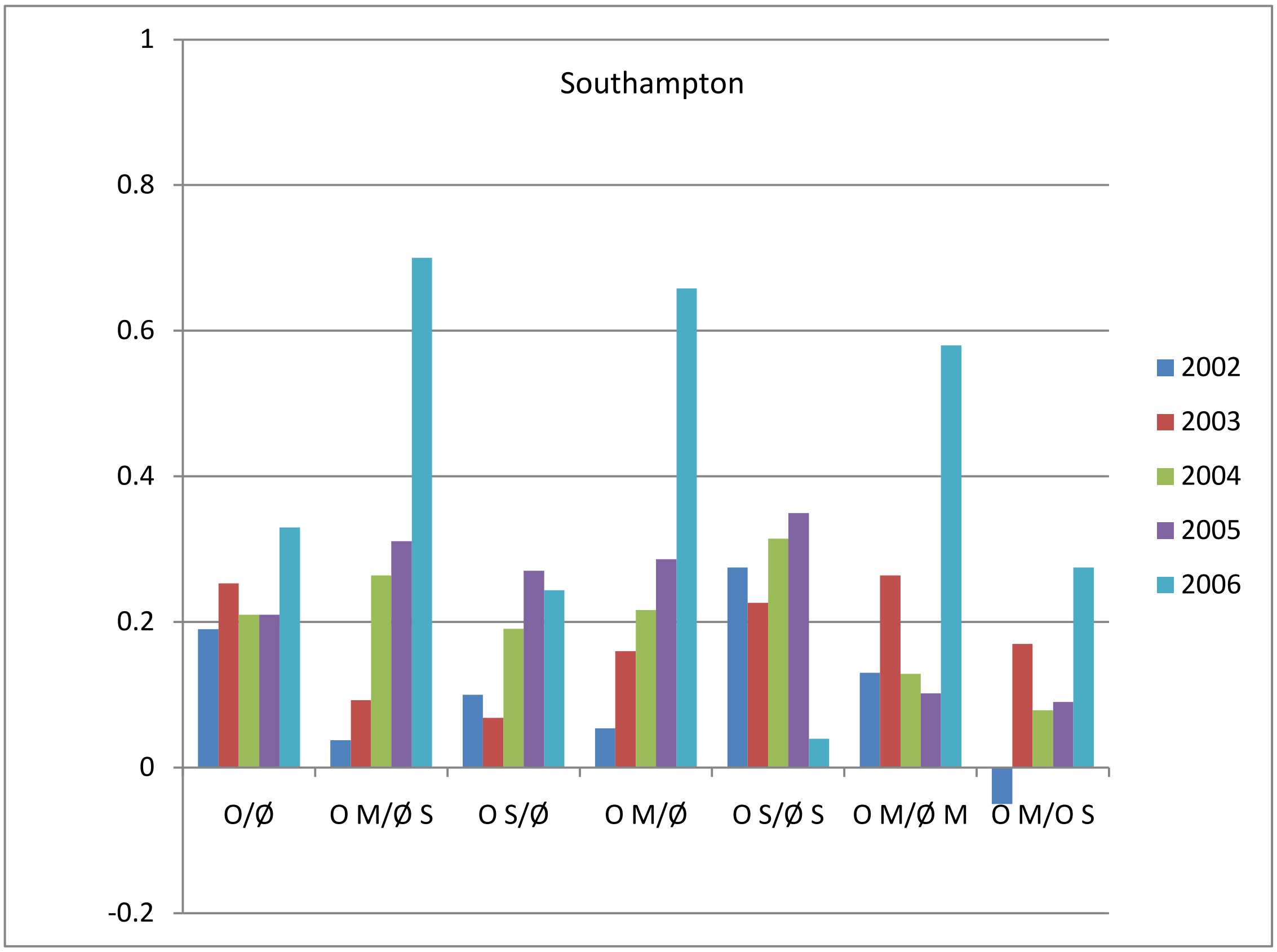

**Figure 12:** ***OA Impact Advantage for Self-Selected vs Mandatory OA for Southampton ECS***

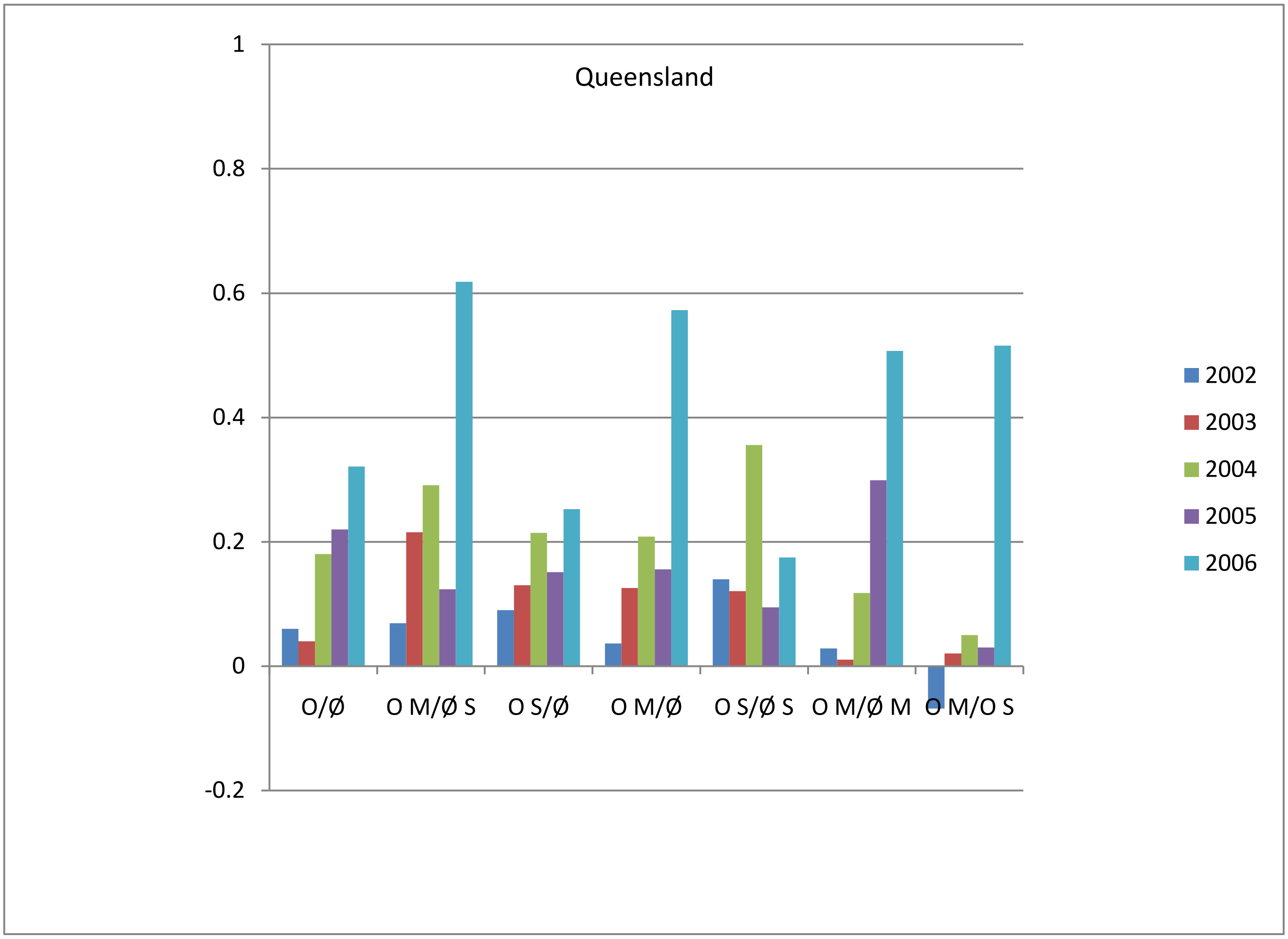

**Figure 13:** ***OA Impact Advantage for Self-Selected vs Mandatory OA for Queensland UT***

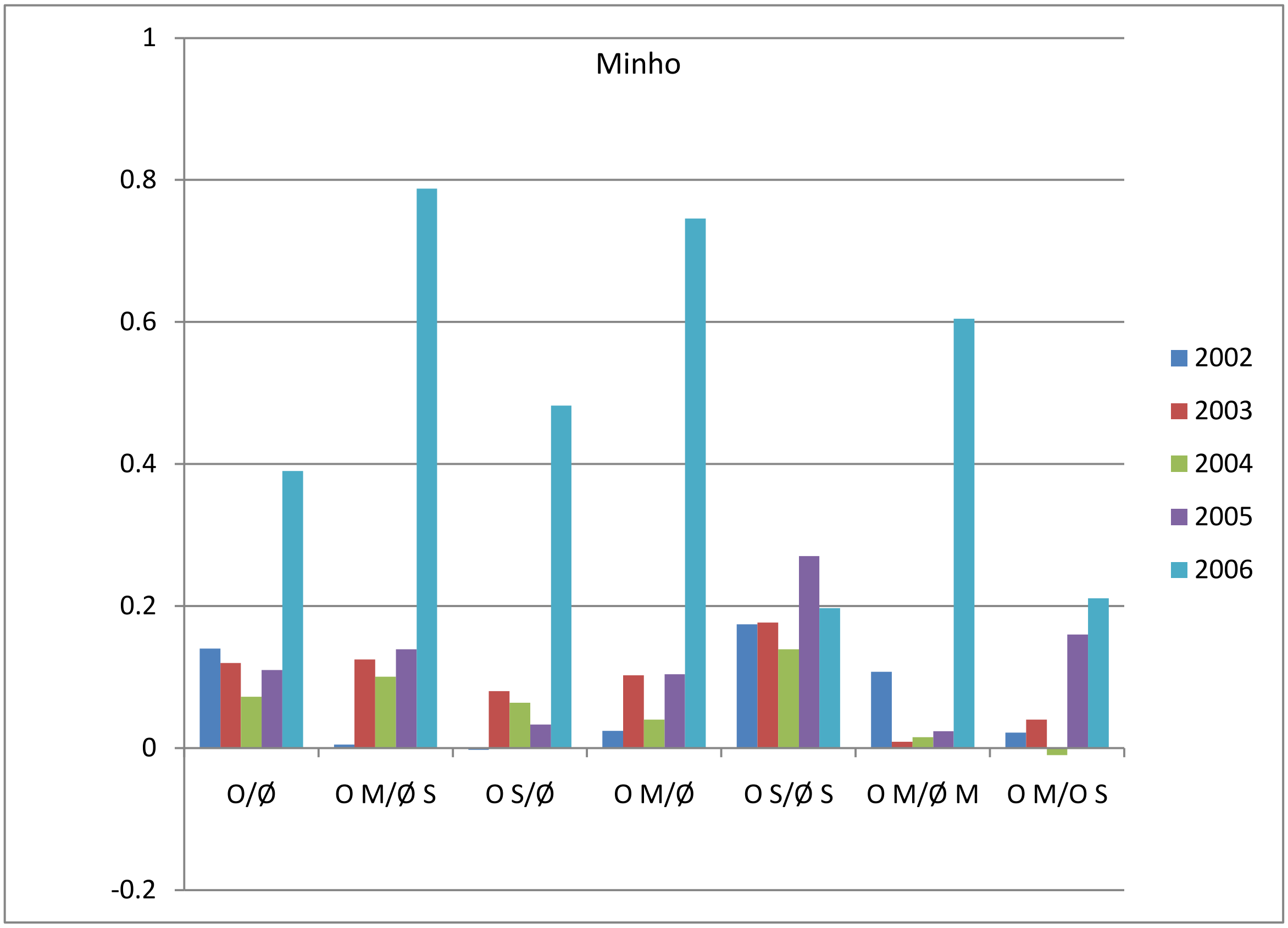

**Figure 14:** ***OA Impact Advantage for Self-Selected vs Mandatory OA for Minho***

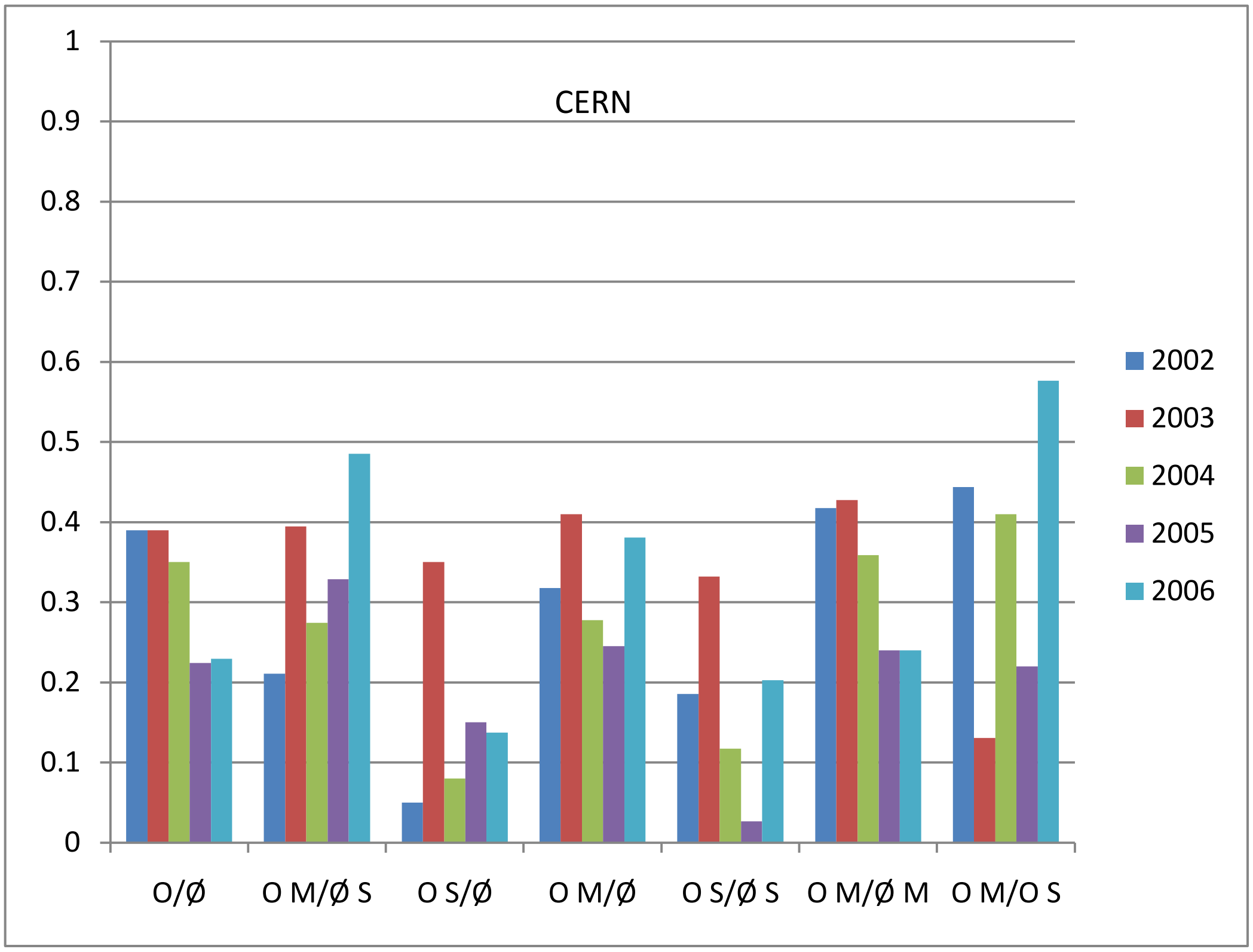

**Figure 15:** ***OA Impact Advantage for Self-Selected vs Mandatory OA for CERN***

**Appendix 2:** ***Multiple regression by JIF – Beta values***

A. JIF1 (JIF < 0,633)

| Model N. | Dependent Var. | Age | Ref_N | Auth_N | Page_N | OA | M | USA | Review | Sci | CERN | South | Minho | Queens | Age*OA |
|---|---|---|---|---|---|---|---|---|---|---|---|---|---|---|---|
| M_1 | Cit_a_0&1-5 | **1,537** | **1,017** | **1,079** | | | | | | | | | **0,701** | | **1,093** |
| M_2 | Cit_a_1-5&5-10 | **1,847** | **1,013** | **1,066** | | | **1,881** | | | | | | | | *1,059* |
| M_3 | Cit_a_1-5&10-20 | **2,071** | **1,026** | *1,054* | **0,962** | *1,533* | **1,902** | | | | | | | | |
| M_4 | Cit_a_1-5&20+ | **2,689** | **1,020** | **1,087** | | *2,406* | | | *4,760* | *3,214* | | | | | |

***Table 4: The Exp(ß) values for logistic regressions for GIF1***

B. JIF2 (0,633 <= JIF < 1,053)

| Model N. | Dependent Var. | Age | Ref_N | Auth_N | Page_N | OA | M | USA | Review | Sci | CERN | South | Minho | Queens | Age*OA |
|---|---|---|---|---|---|---|---|---|---|---|---|---|---|---|---|
| M_1 | Cit_a_0&1-5 | **1,407** | **1,016** | **1,028** | | | ***1,265*** | | *0,605* | | **0,511** | | | | |
| M_2 | Cit_a_1-5&5-10 | **1,548** | **1,012** | | | **1,346** | **1,963** | | | | | | | | |
| M_3 | Cit_a_1-5&10-20 | **1,869** | **1,018** | **1,007** | | *1,337* | **1,722** | | | | | | | | |
| M_4 | Cit_a_1-5&20+ | **2,117** | **1,011** | | | **2,322** | | | **3,106** | | | | | | |

***Table 5: The Exp(ß) values for logistic regressions for GIF2***

C. JIF3 (1,053 <= JIF < 1,743)

| Model N. | Dependent Var. | Age | Ref_N | Auth_N | Page_N | OA | M | USA | Review | Sci | CERN | South | Minho | Queens | Age*OA |
|---|---|---|---|---|---|---|---|---|---|---|---|---|---|---|---|
| M_1 | Cit_a_0&1-5 | **1,581** | **1,012** | **1,032** | | *1,236* | | | | | **0,401** | | | **1,856** | |
| M_2 | Cit_a_1-5&5-10 | **1,540** | **1,007** | **1,033** | | | **1,428** | **1,330** | | | | | | | |
| M_3 | Cit_a_1-5&10-20 | **1,879** | **1,013** | *1,026* | | *1,263* | | | | | | | | *1,382* | |
| M_4 | Cit_a_1-5&20+ | **2,305** | **1,009** | **1,041** | **1,026** | **1,449** | **1,492** | **1,791** | *1,939* | | | **3,734** | | | |

***Table 6: The Exp(ß) values for logistic regressions for GIF3***

D. JIF4 (1,743 <= JIF < 2,468)

| Model N. | Dependent Var. | Age | Ref_N | Auth_N | Page_N | OA | M | USA | Review | Sci | CERN | South | Minho | Queens | Age*OA |
|---|---|---|---|---|---|---|---|---|---|---|---|---|---|---|---|
| M_1 | Cit_a_0&1-5 | **1,690** | **1,020** | | | | | | | *2,090* | *0,554* | *0,233* | | | |
| M_2 | Cit_a_1-5&5-10 | **1,427** | **1,010** | | | | **1,645** | | | | *1,657* | | | | |
| M_3 | Cit_a_1-5&10-20 | **1,800** | **1,019** | | | | **1,729** | | | | **1,615** | | | | |
| M_4 | Cit_a_1-5&20+ | **2,540** | **1,024** | **0,994** | *1,028* | **1,747** | | **1,822** | | | **3,974** | | | | |

***Table 7: The Exp(β) values for logistic regressions for GIF4***

E. JIF5 (2.468 <= JIF < 29,957)

| Model N. | Dependent Var. | Age | Ref_N | Auth_N | Page_N | OA | M | USA | Review | Sci | CERN | South | Minho | Queens | Age*OA |
|---|---|---|---|---|---|---|---|---|---|---|---|---|---|---|---|
| M_1 | Cit_a_0&1-5 | **1,484** | **1,016** | | | | | | **0,182** | | **0,446** | | | | |
| M_2 | Cit_a_1-5&5-10 | **1,312** | **1,010** | | **0,976** | | **1,468** | **1,391** | *0,586* | | | | | | |
| M_3 | Cit_a_1-5&10-20 | **1,590** | **1,007** | 0,*998* | | | | **1,360** | | | | | | **1,751** | |
| M_4 | Cit_a_1-5&20+ | **2,259** | **1,009** | **0,995** | | **1,722** | | **1,635** | *1,650* | *2,007* | | | | | |

***Table 8: The Exp(β) values for logistic regressions for JIF5***